\documentclass[aps,prc,twocolumn,superscriptaddress]{revtex4-1}

\usepackage{graphicx}
\usepackage{amsmath}

\begin{document}


\title{Medium modification of $\gamma$-jet fragmentation functions in Pb+Pb collisions at LHC}


\author{Wei Chen}
\affiliation{School of Nuclear Science and Technology, University of Chinese Academy of Sciences, Beijing 100049, China}
\affiliation{Key Laboratory of Quark and Lepton Physics (MOE) and Institute of Particle Physics, Central China Normal University, Wuhan 430079, China}
\author{Shanshan Cao}
\affiliation{Department of Physics and Astronomy, Wayne State University, Detroit, MI, 48201, USA}
\affiliation{Cyclotron Institute, Texas A\&M University, College Station, TX, 77843, USA}
\author{Tan Luo}
\affiliation{Instituto Galego de F\'isica de Altas Enerx\'ias IGFAE, Universidade de Santiago de Compostela, E-15782 Galicia-Spain}
\author{Long-Gang Pang}
\affiliation{Key Laboratory of Quark and Lepton Physics (MOE) and Institute of Particle Physics, Central China Normal University, Wuhan 430079, China}
\author{Xin-Nian Wang}
\affiliation{Key Laboratory of Quark and Lepton Physics (MOE) and Institute of Particle Physics, Central China Normal University, Wuhan 430079, China}
\affiliation{Nuclear Science Division MS 70R0319, Lawrence Berkeley National Laboratory, Berkeley, California 94720}


\date{\today}

\begin{abstract}
Coupled linear Boltzmann transport and hydrodynamic (CoLBT-hydro) model has been developed for simultaneous simulations of jet propagation and jet-induced medium excitation in heavy-ion collisions. Within this coupled approach, the final reconstructed jets in heavy-ion collisions include not only hadrons from the hadronization of medium modified jet shower partons from the linear Boltzmann transport  (LBT) but also hadrons from the freeze-out of the jet-induced medium excitation in the hydrodynamic evolution of the bulk medium.  Using the CoLBT-hydro model, we study medium modification of the fragmentation functions of $\gamma$-triggered jets in high-energy heavy-ion collisions at the Large Hadron Collider. The CoLBT-hydro model is shown to describe the experimental data not only on the suppression of leading hadrons within the jet cone at large momentum fraction $z_\gamma=p_T^h/p_T^\gamma$ relative to the transverse momentum of the trigger photon due to parton energy loss but also the enhancement of soft hadrons at small $z_\gamma$ and $z_{\rm jet}=p_T^h/p_T^{\rm jet}$ due to jet-induced medium excitation. There is no suppression of the fragmentation function, however, at large momentum fraction $z_{\rm jet}$ relative to the transverse momentum of the reconstructed jet due to trigger bias and medium modification of quark to gluon jet fraction.  For jets whose final transverse momenta are comparable to or larger than that of the trigger photon, the trigger bias can lead to enhancement of the jet fragmentation function at large $z_{\rm jet}$.

\end{abstract}


\maketitle


\section{ Introduction }
Experimental data at the Relativistic Heavy-Ion Collider (RHIC) have provided solid evidences for  the formation of the strongly coupled quark-gluon plasma (QGP) in high-energy heavy-ion collisions \cite{Adams:2005dq,Adcox:2004mh,Gyulassy:2004zy}.  Exploration and extraction of transport properties of QGP at different energy scales through hard and soft probes have become the current focus of theoretical and experimental studies of heavy-ion collisions at both RHIC and the Large Hadron Collider (LHC). One of the hard probes that one can use to study QGP properties is jet tomography \cite{Gyulassy:2001nm,Wang:2002ri}. Such jet tomography is based on the early idea that parton energy loss due to interaction with the QGP medium can lead to suppression of high-energy jets and hadrons in heavy-ion collisions relative to elementary proton-proton collisions at the same colliding energy. This phenomenon is often referred  to as  ``jet quenching" \cite{Gyulassy:1990ye,Wang:1991xy}. Theoretical calculations \cite{Bjorken:1982tu,Baier:1996kr,Baier:1998kq,Zakharov:1996fv,Gyulassy:2000fs,Gyulassy:2000er,Wiedemann:2000za, Arnold:2002ja, Guo:2000nz,Wang:2001ifa,Gyulassy:2003mc}  show that parton energy loss is directly related to the jet transport coefficient of the dense medium and one can extract the jet transport coefficient through phenomenological study of experimental data on jet quenching \cite{Burke:2013yra}.

Jet medium interaction is also shown to lead to medium modification of full jet production rate and jet substructures in high-energy heavy-ion collisions \cite{Vitev:2009rd,Aad:2010bu,Qin:2010mn,Young:2011qx,Dai:2012am,Chien:2015hda,Milhano:2015mng,
Chang:2016gjp,Blaizot:2015lma,Kang:2017frl,Neufeld:2012df,He:2011pd,Renk:2012cx,Chang:2016gjp}. Full jets are reconstructed from collimated clusters of hadrons within a given jet-cone in experimental measurements. In heavy-ion collisions, however, final jets are not only modified by energy loss of leading partons through both elastic and inelastic collisions but are also influenced by the redistribution of the lost energy in the form of radiated gluons which must go through additional rescattering \cite{Wang:2013cia,Blaizot:2013hx,Apolinario:2012cg,Casalderrey-Solana:2015vaa} and jet-induced medium response \cite{Tachibana:2017syd,Chen:2017zte,Cao:2020wlm}. 
In the transport description of jet propagation in the QGP medium,  jet-induced medium response is the result of  the transport of recoil medium partons from jet-medium interaction.
 Part of the final hadrons from this jet-induced medium response will fall into the jet-cone and be considered as part of the jet. 
 These hadrons from recoil partons in jet-induced medium response will then contribute to the total energy within the jet-cone and affect the medium modification of the full jet production rate. Effects of jet-induced medium response have been clearly illustrated in the calculation of medium modification of the single inclusive jet production rate \cite{He:2018xjv}  and $\gamma/Z^0$-jet correlations \cite{Luo:2018pto,Zhang:2018urd} within the Linear Boltzmann Transport (LBT) model \cite{Wang:2013cia,Li:2010ts,He:2015pra,Cao:2016gvr,Cao:2017hhk}.
They should also affect the distribution of particles within the jet-cone and lead to some unique modification of the jet substructures such as the jet fragmentation functions \cite{Chen:2017zte} and transverse jet profile \cite{Tachibana:2017syd}.

Within the LBT model, both jet shower and recoil partons are assumed to interact with the QGP medium according to perturbative QCD. This assumption becomes problematic for low energy shower and recoil partons whose interaction with the medium could become non-perturbative. If the number density of recoil partons becomes comparable to the bulk medium parton density, interaction among recoil partons can become important which is neglected in LBT.  In the extreme limit of the strong interaction between recoil and medium partons, one can assume that all recoil and radiated partons become thermalized in the medium. Their further transport in the medium can be described by hydrodynamics \cite{Stoecker:2004qu,Casalderrey-Solana:2006zdq,Betz:2008ka,Neufeld:2009ep,Qin:2009uh,Bouras:2014rea,Tachibana:2017syd} with a source term provided by the energy and momentum that is deposited by the propagating jet.  The recently  developed coupled LBT and hydrodynamic (CoLBT-hydro) model  \cite{Chen:2017zte}  takes a middle approach in which only soft radiated and recoil partons below an energy scale are included in the source term for a viscous hydrodynamics while transport of energetic partons from induced radiation and recoil is described by LBT. Furthermore, both the bulk medium and the source term are updated simultaneously in LBT and CLVisc hydrodynamics in real time, therefore giving the name CoLBT-hydro. It has been applied to describe the experimental data on $\gamma$-hadron correlations in heavy-ion collisions  \cite{Chen:2017zte}.

In this paper, we will employ the CoLBT-hydro model to study medium modification of fragmentation functions of $\gamma$-triggered jets in high-energy heavy-ion collisions at LHC. We will specifically look at the influence of jet-induced medium response on the jet fragmentation functions. We will start with a brief introduction of the CoLBT-hydro model and discuss about the constraints on the initial conditions for the CLVisc hydrodynamics by the bulk hadron spectra.  We will then report the CoLBT-hydro results on jet fragmentation functions in p+p and its medium modification $I_{AA}$ in central and peripheral Pb+Pb collisions at $\sqrt{s_{\rm NN}}$=5.02 TeV as compared to data from CMS and ATLAS experiments at LHC.

\section{ CoLBT-hydro model }

In the CoLBT-hydro model~\cite{Chen:2017zte}, jet propagation within LBT is coupled to the dynamic evolution of the bulk medium according to a (3+1)D relativistic hydrodynamic model  in real time. In this coupled approach, the energy and momentum lost by jet shower and recoil partons in each time step  is transferred to the bulk medium through a source term in the hydrodynamic equations which in turn update the local temperature and fluid velocity of the  bulk medium for the transport of jet shower and recoil partons in the next time step. It essentially combines the pQCD approach for the propagation of energetic jet shower and recoil partons with the hydrodynamic evolution of the strongly coupled QGP medium with real time coupling between the two.

The LBT model \cite{Wang:2013cia,Li:2010ts,He:2015pra,Cao:2016gvr,Cao:2017hhk} is developed for  jet propagation and transport in QGP with an emphasis on the transport of both jet shower and medium recoil partons on an equal footing.  The basic building block of the LBT model is the interaction probability within a given time step of the parton propagation,
\begin{equation}
P^{a}_{\rm tot} = P^a_{\rm el}(1-P^a_{\rm inel})+P^a_{\rm inel},
\end{equation}
which can be separated into the probability for pure elastic scattering (first term) and that for inelastic scattering with at least one gluon radiation (the second term), where
\begin{equation}
P^a_{\rm el} = 1-\exp[-\Delta \tau \Gamma^{\rm el}_{a}(x)],
\end{equation}
and
\begin{equation}
P^a_{\rm inel} = 1-\exp[-\Delta \tau \Gamma^{\rm inel}_{a}(x)]
\end{equation}
are the probabilities for at least one elastic scattering and an inelastic process in a time step $\Delta\tau$ during the propagation of parton $a$ at the location $x$, respectively.  Given the local medium information such as the temperature $T$, parton density $\rho_a$ and fluid velocity $u$, the elastic scattering rate $\Gamma^{\rm el}_{a}$ for parton $a$ is
\begin{equation}
\Gamma^{\rm el}_{a} = \frac{p \cdot u}{p_0}\sum_{bcd}\rho_b(x) \sigma_{ab\rightarrow cd},
\end{equation}
where the summation is over all possible parton flavors
and channels of scattering with the cross section  $\sigma_{ab\rightarrow cd}$. The gluon radiation rate $\Gamma^{\rm inel}_{a}$ is given by
\begin{equation}
\Gamma_a^{\rm inel}=\frac{1}{1+\delta_{ag}}\int dz dk_\perp^2 \frac{dN^a_g}{dzdk_\perp^2d\tau},
\end{equation}
where $\delta_{ag}$ is the Kronecker delta function to account for the degeneracy of the final state for $g\rightarrow gg$ splitting.
The differential gluon radiative spectra  is assumed to follow that from the high-twist approach \cite{Guo:2000nz,Wang:2001ifa,Zhang:2003wk,Zhang:2004qm}
\begin{equation}
\frac{dN^a_g}{dzdk_\perp^2d\tau}
=\frac{6\alpha_{s}P_{a}(z) k^{4}_{\perp}}{\pi (k^{2}_{\perp}+z^{2}m^{2})^{4}}\cdot \frac{p\cdot u}{p_{0}}\hat{q}_{a}(x)\sin^{2}\left (\frac{\tau-\tau_{i}}{2\tau_{f}}\right ),
\end{equation}
where $P_{a}(z)$ is the splitting function for the propagating parton $a$ to emit a gluon with the energy fraction $z$ and transverse momentum $k_{\perp}$, $m$ is the mass of the propagating parton, $\tau_f=2p^0 z(1-z)/(k^2_{\perp}+z^2 m^2)$ is the gluon formation time, $\tau_i$ is the time of the last gluon emission, and the jet transport parameter,
\begin{equation}
\hat{q}_{a}(x)=\sum_{bcd}\rho_{b}(x)\int d\hat t q_\perp^2 \frac{d\sigma_{ab\rightarrow cd}}{d\hat t},
\end{equation}
is defined as the transverse momentum transfer squared per unit length in the local comoving frame. Note that the parton density  $\rho_{b}(x)$ here includes the degeneracy factor.  We refer readers to Refs. \cite{Wang:2013cia,Li:2010ts,He:2015pra,Cao:2016gvr,Cao:2017hhk} for more details about the LBT model.

In order to simulate parton transport concurrently with a relativistic hydrodynamic model which is normally formulated in the Milne coordinates $(\tau,x,y,\eta_{s})$, LBT is also reformulated in the same Milne coordinates in the CoLBT-hydro model,  where $\tau = \sqrt{t^{2}-z^{2}}$ and $\eta_{s}= (1/2) \ln[(t+z)/(t-z)]$ are the proper time and the space-time rapidity in terms of the Cartesian coordinates $(t,x,y,z)$. The CoLBT-hydro model employs the CCNU-LBNL viscous (CLVisc) (3+1)D hydrodynamic  model \cite{Pang:2012he,Pang:2018zzo} to solve  the hydrodynamic equations,
\begin{equation}
\partial_{\mu}T^{\mu\nu}=J^{\nu},
\label{eq:hydro}
\end{equation}
for the bulk medium evolution concurrently with LBT with a source term $J^{\nu}$ updated in real  time and a parametrized equation of state (EoS) s95p-v1~\cite{Huovinen:2009yb}. An energy cut-off $p^0_{\rm cut}$ of parton energy in the comoving frame of the fluid cell is introduced to separate soft and hard partons in CoLBT-hydro. For hard partons with the energy $p\cdot u>p^0_{\rm cut}$, their transport through the medium is simulated according to LBT.  Soft partons with the energy $p\cdot u<p^0_{\rm cut}$ are, however, assumed to become thermalized with the medium and their energy and momentum will be deposited into the medium as the source term in the hydrodynamic equations. In LBT, initial medium partons that participate in the jet-medium interaction are subtracted from the final state as ``negative" partons according to the back-reaction in the Boltzmann transport equations.  The energy and momentum of these ``negative" partons ($p\cdot u<0$) are also included in the source term in order to ensure energy-momentum conservation in CoLBT-hydro. With the above division between hard parton transport and soft parton evolution, the source term in CoLBT-hydro can be expressed as,
\begin{equation}
\begin{aligned}
J^{\nu}=\sum_{i} \frac{\theta(p^{0}_{\rm cut}-p_{i}\cdot u)p^{\nu}}{\tau(2\pi)^{3/2}\sigma^{2}_{r}\sigma_{\eta_{s}}\Delta\tau}
e^{-\frac{(\vec{x}_{\perp}-\vec{x}_{\perp i})^{2}}{2\sigma^{2}_{r}}
-\frac{(\eta_{s}-\eta_{si})^{2}}{2\sigma^{2}_{\eta_{s}}}}
\end{aligned},
\label{eq:smear}
\end{equation}
with a Gaussian smearing in the Milne coordinates, where the summation is over all jet shower, medium recoil and ``negative" partons. The Gaussian half widths $\sigma_r$=0.2~fm, $\sigma_{\eta_s}$=0.2 are assumed.  The energy cut-off  $p^0_{\rm cut}$ in principle can depend on the local temperature. We will however consider it a constant with a value of 2.0 GeV/$c$ unless specifically stated. Here we have assumed instantaneous local thermalization of the deposited energy and momentum and neglect the causality violations whose effects should be negligible \cite{Tachibana:2020atn}. 

For each time step in the LBT transport model, the source term will be calculated according to Eq.~(\ref{eq:smear}) with which bulk medium evolution at the next time $\tau+\Delta \tau$ will be evaluated according to the hydrodynamic equations in Eq.~(\ref{eq:hydro}). Hard partons will propagate according to the LBT model in the  updated bulk medium along their classical trajectories in the time step $\tau +\Delta \tau$. 
This coupled LBT parton transport and hydrodynamic evolution of the bulk medium are iterated until the end of hydrodynamic evolution.  

\begin{figure}
\includegraphics[width=3.5in]{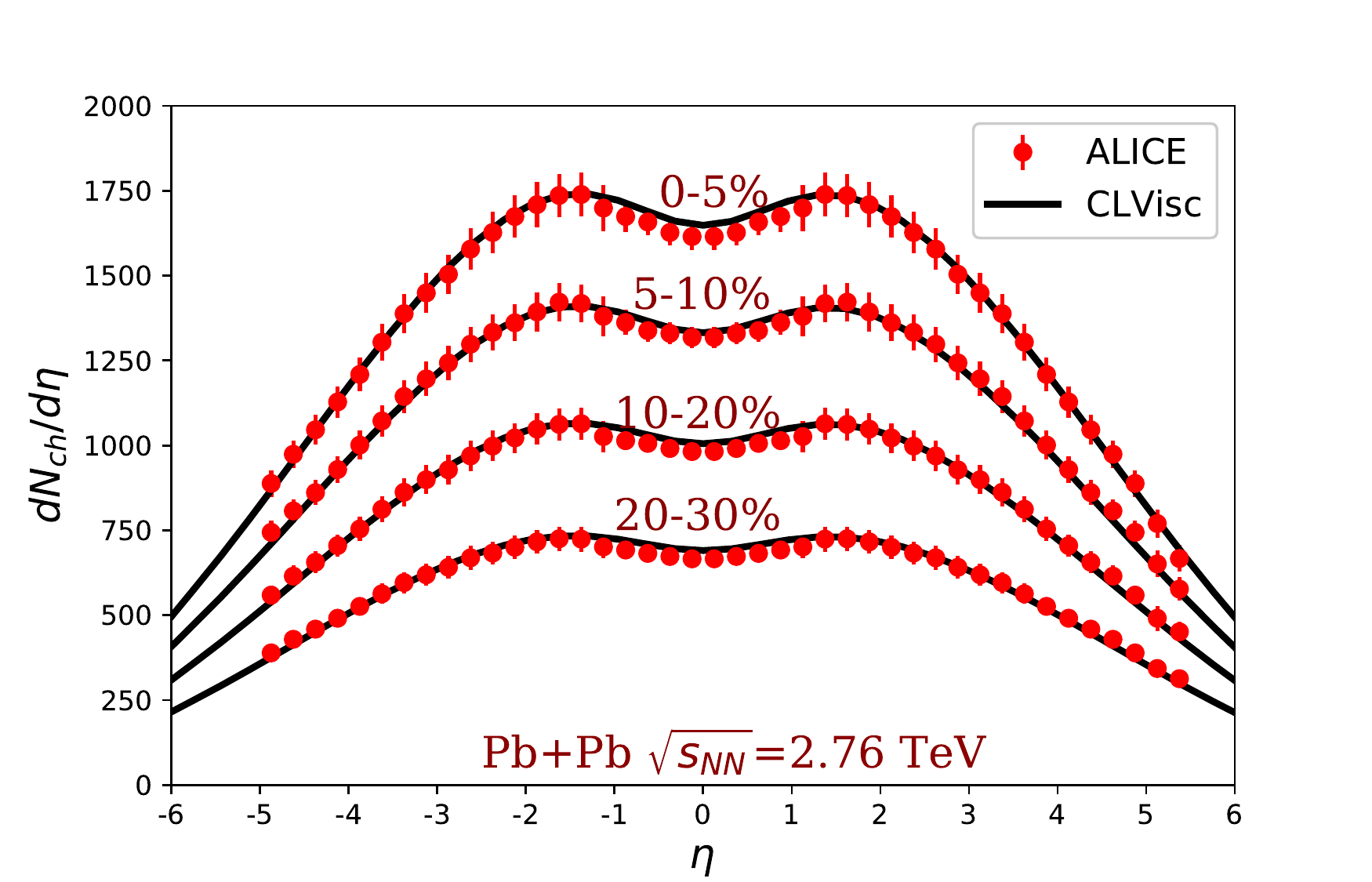}
\includegraphics[width=3.5in]{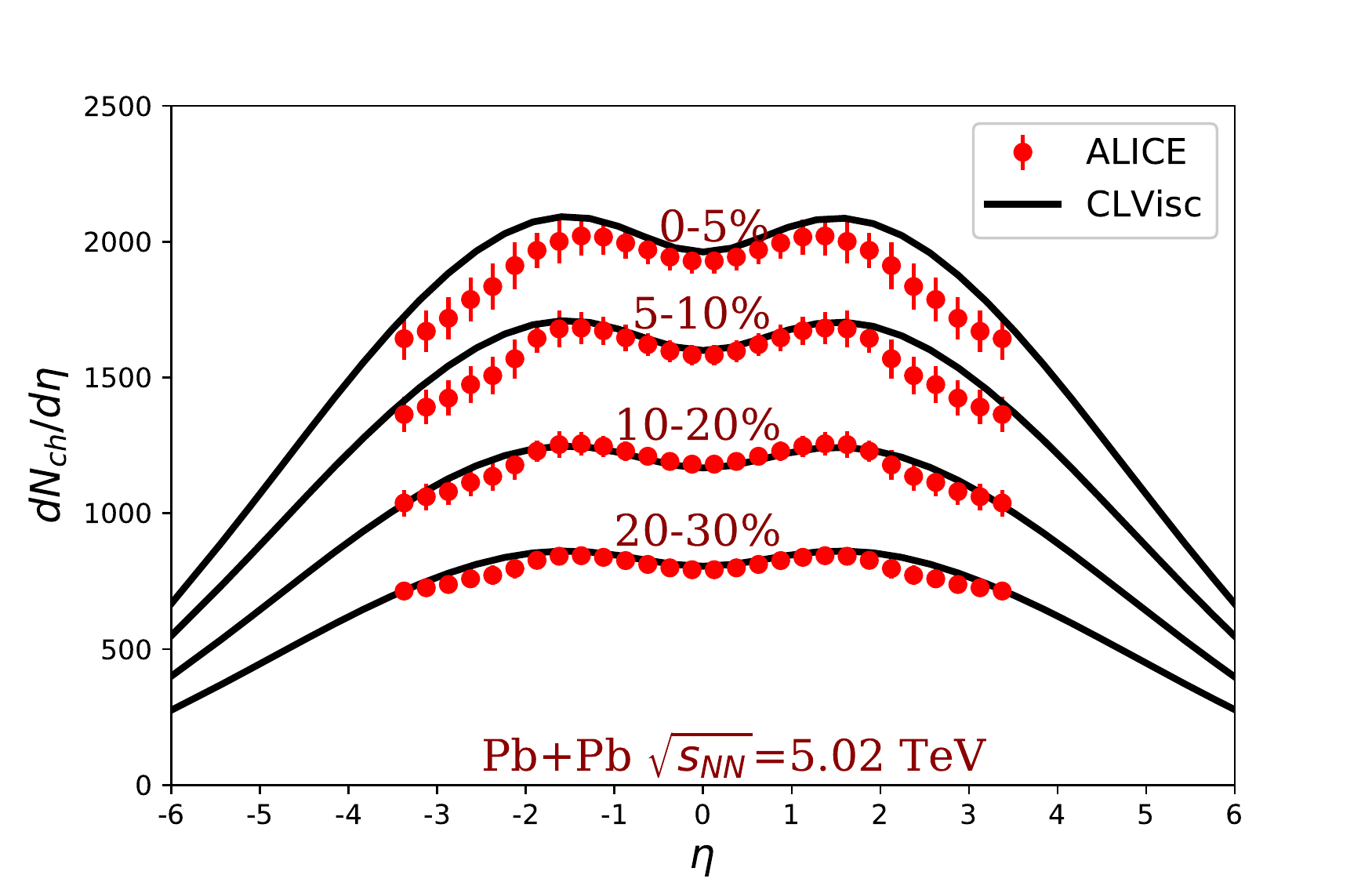}
\caption{Charged hadron pseudo-rapidity distributions for     Pb+Pb collisions at $\sqrt{s_{\rm NN}}$ = 2.76 TeV and 5.02 TeV from event-by-event CLVisc hydrodynamic simulations compared to experimental data~\cite{Abbas:2013bpa,Adam:2016ddh}. The centrality classes of heavy-ion collisions are defined according to the initial parton multiplicity distribution from AMPT model~\cite{Lin:2004en} in our simulations. }
\label{fig-dndeta}
\end{figure}

The initial energy-momentum density distributions for event-by-event CoLBT-hydro simulations are obtained from particles in A Multi-Phase Transport (AMPT) model simulations \citep{Lin:2004en}. Both transverse and longitudinal fluctuations are taken into account event-by-event.  The normalization factor of the initial energy-momentum density, the initial time $\tau_0$ = 0.2 fm/$c$ and freeze-out temperature $T_{\rm f}$ = 137 MeV are adjusted to reproduce the experimental data on the final charged hadron rapidity distribution in Pb+Pb collisions at $\sqrt{s_{\rm NN}}$ = 2.76 and 5.02 TeV~\cite{Abbas:2013bpa,Adam:2016ddh}, as shown in Fig.~\ref{fig-dndeta}. 

We employ the parton recombination model \cite{Han:2016uhh} developed within the JET Collaboration for the hadronization of hard partons from the LBT transport. Hadron spectra from jet-induced medium excitation are calculated as the difference between the bulk hadron spectra from CLVisc via Cooper-Frye \cite{Cooper:1974mv} freeze-out of the bulk medium with and without the source term induced by jets. The final hadron spectra from CoLBT-hydro include contributions from both hadronization of hard jet shower partons in LBT and soft hadrons from jet-induced medium excitation.

\begin{figure*}[htbp]
\includegraphics[width=3.25in]{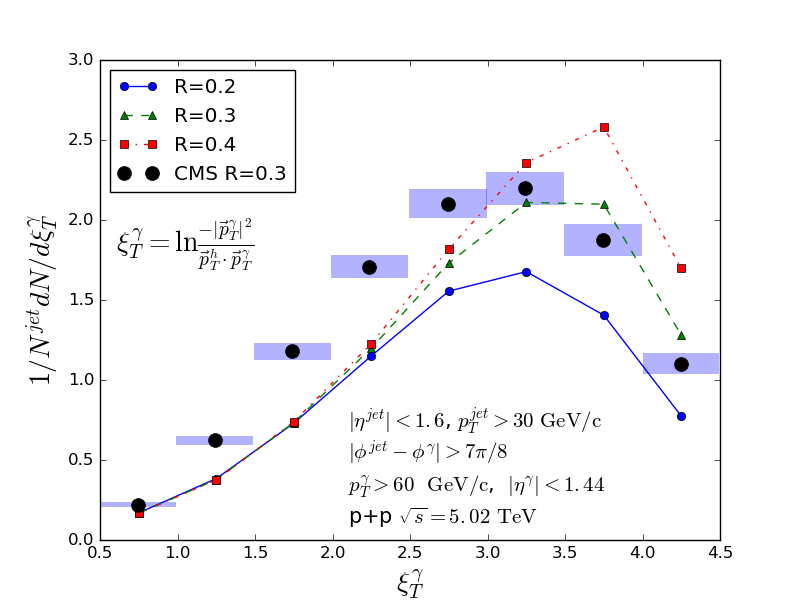}
\includegraphics[width=3.25in]{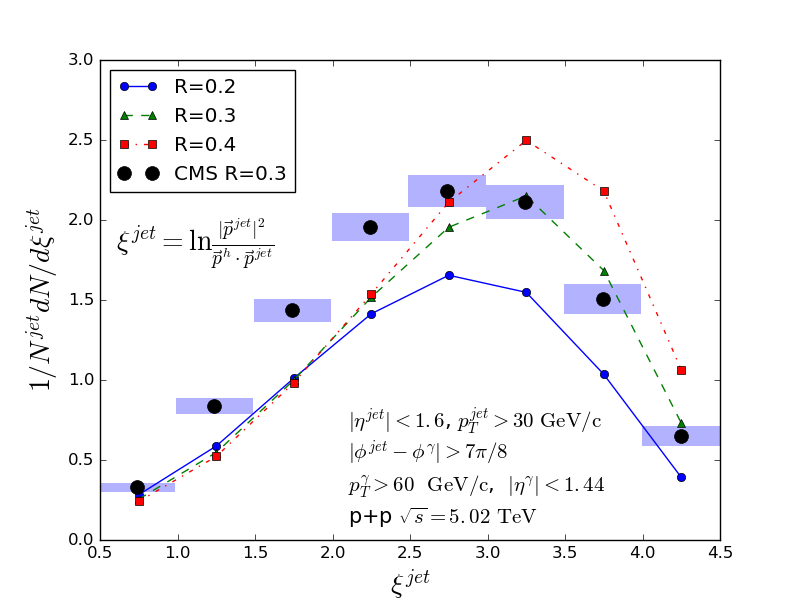}
\caption{$\gamma$-jet fragmentation function as a function of $\xi^{\gamma}_{T}$ and $\xi^{\rm jet}$ in p+p collisions at $\sqrt{s}=5.02$ TeV for different jet-cone sizes as compared to CMS data. The recombination model is used in the hadronization processes.}
\label{fig-pp-recom}
\end{figure*}

\begin{figure*}[htbp]
\includegraphics[width=3.25in]{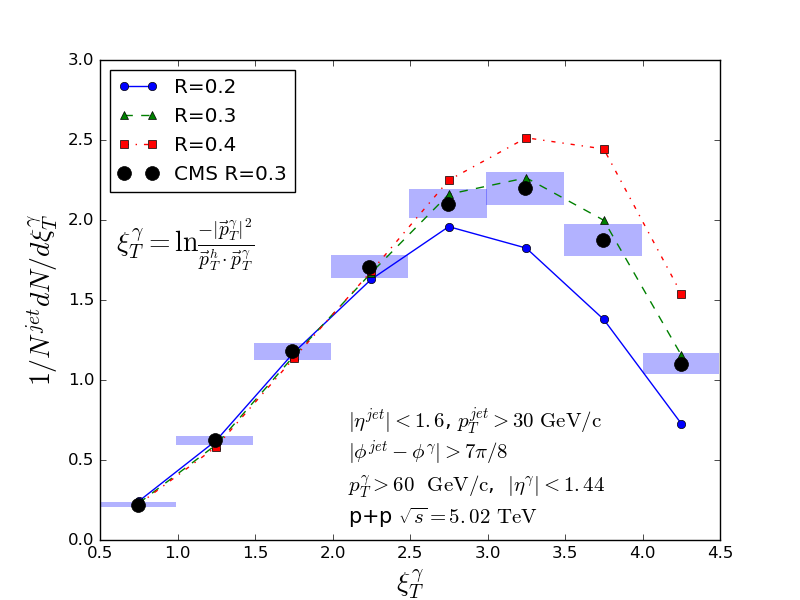}
\includegraphics[width=3.25in]{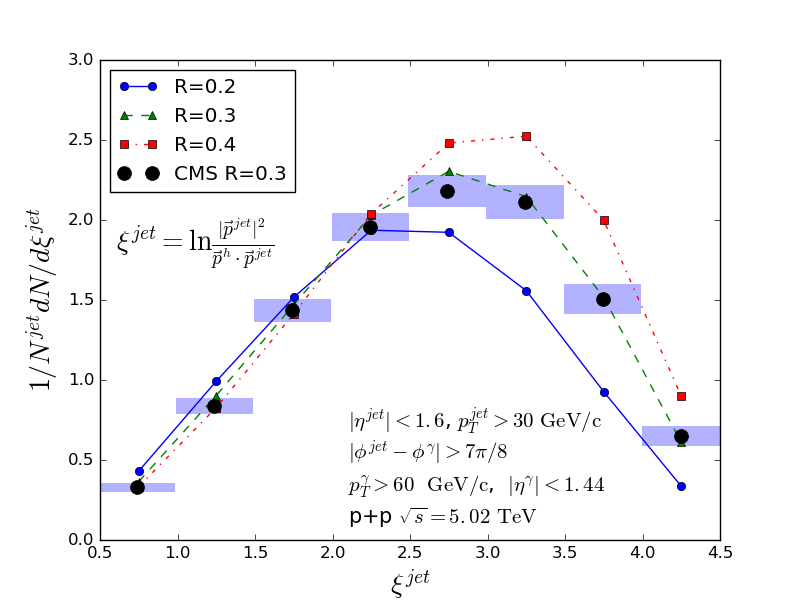}
\vspace{0.2cm}
\caption{$\gamma$-jet fragmentation function as a function of $\xi^{\gamma}_{T}$ and $\xi^{\rm jet}$ in p+p collisions at $\sqrt{s}=5.02$ TeV for different jet-cone sizes as compared to CMS data. Pythia8 is used for hadronization processes which include string fragmentation and secondary decays.}
\label{fig-pp-pythia}
\end{figure*}

\section{ $\gamma$-jet fragmentation function in p+p collisions}

Since photons do not participate in the strong interaction with the QGP  medium, they can provide the best calibration of the transverse energy and direction of the initial hard partons in $\gamma$-jet processes. One can then best study the medium modification of $\gamma$-jet fragmentation function and parton energy loss in QGP in heavy-ion collisions. In our first work with the CoLBT-hydro model \cite{Chen:2017zte}, we carried out a study of the medium modification of $\gamma$-hadron correlations in heavy-ion collisions at RHIC. CoLBT-hydro describes well the suppression of leading hadrons in $\gamma$-hadron correlation due to parton energy loss and predicts an enhancement of soft hadrons due to jet-induced medium excitation. We will calculate in the following the medium modification of $\gamma$-jet fragmentation functions in heavy-ion collisions at LHC.

In this study, we use Pythia8 \cite{Sjostrand:2014zea} to generate initial jet shower partons for $\gamma$-jet events in p+p collisions.  Trigger photons in $\gamma$-jet events are selected according to the same kinematic cuts as in the experiments \cite{Sirunyan:2018qec} to which we will compare:  The transverse momentum is restricted to the range $p_{T}^{\gamma}> 60$  GeV/$c$ and the pseudo-rapidity range to $|\eta^{\gamma}|<1.44$.  In p+p collisions, all final-state particles from jet. shower partons in Pythia8 are used for the jet reconstruction using FASTJET  \cite{Cacciari:2008gp} with the anti-$k_{T}$ algorithm and jet zone size  $R=0.3$. 

In order to ensure that the hadronization mechanisms in both p+p and Pb+Pb collisions are consistent, a parton recombination model developed by the Texas A\&M University group within the JET Collaboration is used for hadronization of both hard jet shower and soft recoil medium partons. In our simulations for p+p collisions, the final partons generated from Pythia8 are used as input to the recombination model for hadronization processes, and jet reconstruction is carried out at the parton level in this case. Reconstructed jets with $|\eta^{\rm jet}|<1.6$ and $p_{T}^{\rm jet}>30$ GeV/$c$ are selected for the analysis. The azimuthal angles between trigger photons and reconstructed jets are restricted to $\Delta \phi_{j\gamma}=|\phi^{\rm jet}-\phi^{\gamma}|>7\pi/8$ as in the experimental analysis. Since the decay processes of neutral particles, such as $\pi^0$ and $K_{s}$, are not taken into account in the recombination model, and final hadrons obtained from the recombination model do not distinguish between charged and neutral particles, we empirically assume that charged particles account for 2/3 of the total number of final hadrons.

Shown in Fig.~\ref{fig-pp-recom} are the $\gamma$-jet fragmentation functions as a function of 
\begin{equation}
\xi^{\gamma}_{T}=\ln(-{p_T^\gamma}^2/\vec p_T^{\,h}\cdot \vec p_T^{\,\gamma}),
\end{equation}
and
\begin{equation}
\xi^{\rm jet}=\ln({p_T^{\rm jet}}^2/\vec p_T^{\,h}\cdot \vec p_T^{\,\rm jet})
\end{equation}
in p+p collisions at $\sqrt{s}=5.02$ TeV for different jet cone sizes ($R=\sqrt{\Delta \eta^{2} + \Delta \phi^{2}}=0.2, 0.3,0.4$) as compared to the CMS data for $R=0.3$ \cite{Sirunyan:2018qec}, where $\vec{p}_T^{\,\rm jet}$, $\vec{p}_T^{\,\gamma}$ and $\vec{p}_T^{\,h}$ are the transverse momenta of the reconstructed jets, direct photon and charged hadrons, respectively. The $\gamma$-jet fragmentation functions obtained with the recombination model are overall all consistent with the experimental data. They, however, underestimate the experimental data  in the intermediate $\xi$ range ($1.25<\xi<3$).

As an alternative to the parton recombination model for hadronization,  we use the hadronic information generated by Pythia8, in which hadronization processes include string fragmentation and secondary and neutral hadron decays, to calculate the fragmentation function of $\gamma$-jet in p+p collisions under the same conditions for jet reconstruction. The results as shown in Fig.~\ref{fig-pp-pythia} are in very good agreement with the experimental data from CMS.  We assume the difference in the jet fragmentation functions between Pythia8 and the recombination model for parton hadronization is the same in p+p and Pb+Pb collisions.  Under this approximation, the ratio between the fragmentation functions from Pythia8 and the recombination model will be applied as a multiplicative factor  to the hadron spectra from CoLBT-hydro with the recombination model to obtain the final jet fragmentation functions in Pb+Pb collisions.

\section{ $\gamma$-jet fragmentation function in Pb+Pb collisions}

In simulations of $\gamma$-jet in CoLBT-hydro model for Pb-Pb collisions at $\sqrt{s_{\rm NN}}=5.02$ TeV, we set the effective strong coupling constant at $\alpha_{\rm s}$=0.16, which is the only parameter that controls parton energy loss in LBT.  The initial positions of the $\gamma$-jet are sampled according to the spatial distribution of binary hard processes from the same AMPT event that provides the initial condition for the hydrodynamic evolution of the bulk medium. These jet partons will propagate through the QGP medium and their lost energy will be transported in the QGP medium in the CoLBT-hydro model after their formation time $\tau_f=2p^0/p^2_T$ or the QGP formation time $\tau_0$ whichever later.

In the jet reconstruction and calculation of jet fragmentation functions, we will include both hadrons from the hadronization of hard jet shower and medium recoil partons in LBT and soft hadrons from the jet-induced medium excitation (j.i.m.e.) from CLVisc hydrodynamics. The specific steps to calculate the jet fragmentation functions in Pb+Pb collisions are as follows:
\begin{itemize}
\item In each CoLBT-hydro simulation of a $\gamma$-jet event in Pb+Pb collisions, we use the final hard partons from LBT to reconstruct jets using FASTJET  \cite{Cacciari:2008gp}with anti-$k_T$ algorithm. Their transverse momenta are denoted as $p_T^{\rm LBT}$. 

\item For each reconstructed jet, we calculate the contribution to the jet's transverse momentum from the jet-induced medium excitation by integrating the final hadron spectra of jet-induced medium excitation in CLVisc within the jet-cone. We neglect the fluctuation of the jet-induced medium excitation within the jet cone so that inclusion of  j.i.m.e. in jet reconstruction will not change the jet direction $(y^{\rm jet},\phi^{\rm jet})$. The final jet transverse momentum is then $p_{T}^{\rm jet}=p_{T}^{\rm LBT}+p_{T}^{\rm j.i.m.e.}$. 

\item With the information of final charged particles from LBT and the hadron spectra from jet-induced medium excitation within the jet zone,
\begin{eqnarray}
\frac{dN}{d\xi^{\gamma}_T}=&\int \frac{dN}{dydp_T d\phi}\delta(\xi^{\gamma}_T-\ln \frac{-|\vec{p}^{\,\gamma}_T|^2}{\vec{p}_T^{\,h}\cdot \vec{p}^{\,\gamma}_T}) dp_T dyd\phi \nonumber\\
&\times \theta(R-\sqrt{(y-y^{\rm jet})^2+(\phi-\phi^{\rm jet})^2}) ,
\end{eqnarray}
one can calculate the hadron distribution or jet fragmentation function as a function of $\xi^{\gamma}_{T}$  within the jet cone.  One can similarly calculate 
 the jet fragmentation function as a function of $\xi^{\rm jet}$.
\end{itemize}


\begin{figure}
\centering
\includegraphics[scale=0.4]{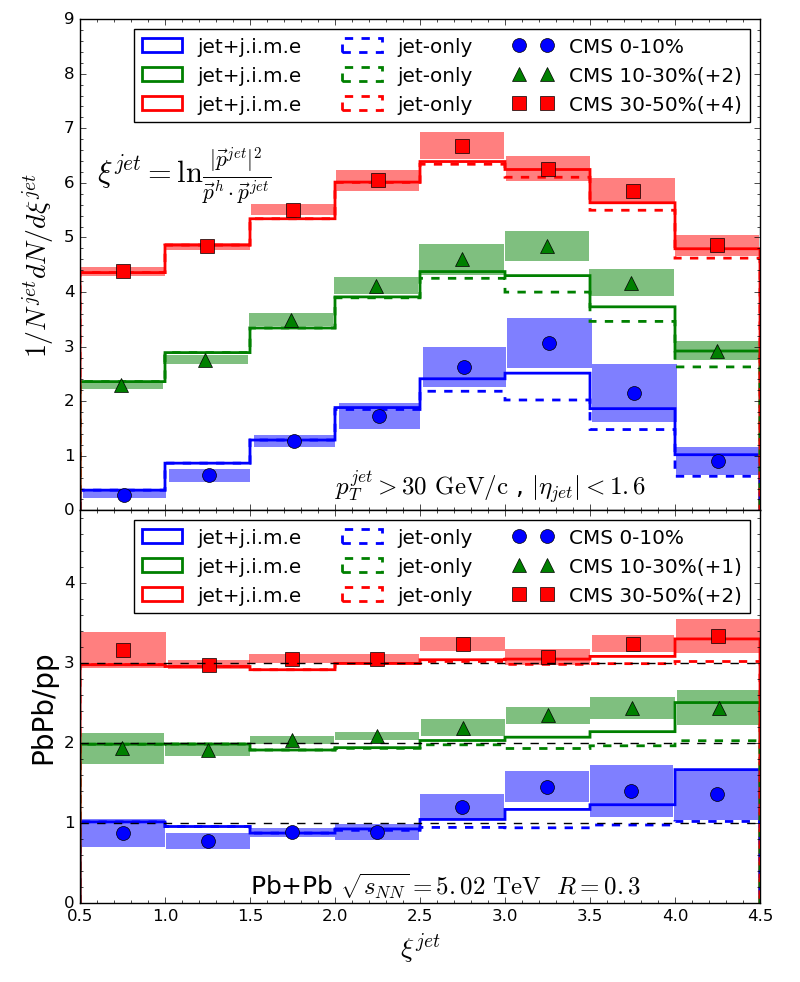} 
\caption{$\gamma$-jet fragmentation function as a function of $\xi^{jet}$ in Pb+Pb collisions at $\sqrt{s_{NN}}$=5.02 TeV for different centrality classes (upper panel) and the corresponding ratio of the Pb+Pb to p+p results (lower panel) as compared to CMS data~\cite{Sirunyan:2018qec}. The solid (dashed) histograms are CoLBT-hydro results with (without) jet-induced medium excitations.}
\label{fig-ratio-jet}
\end{figure}

\begin{figure}
\centering
\includegraphics[scale=0.4]{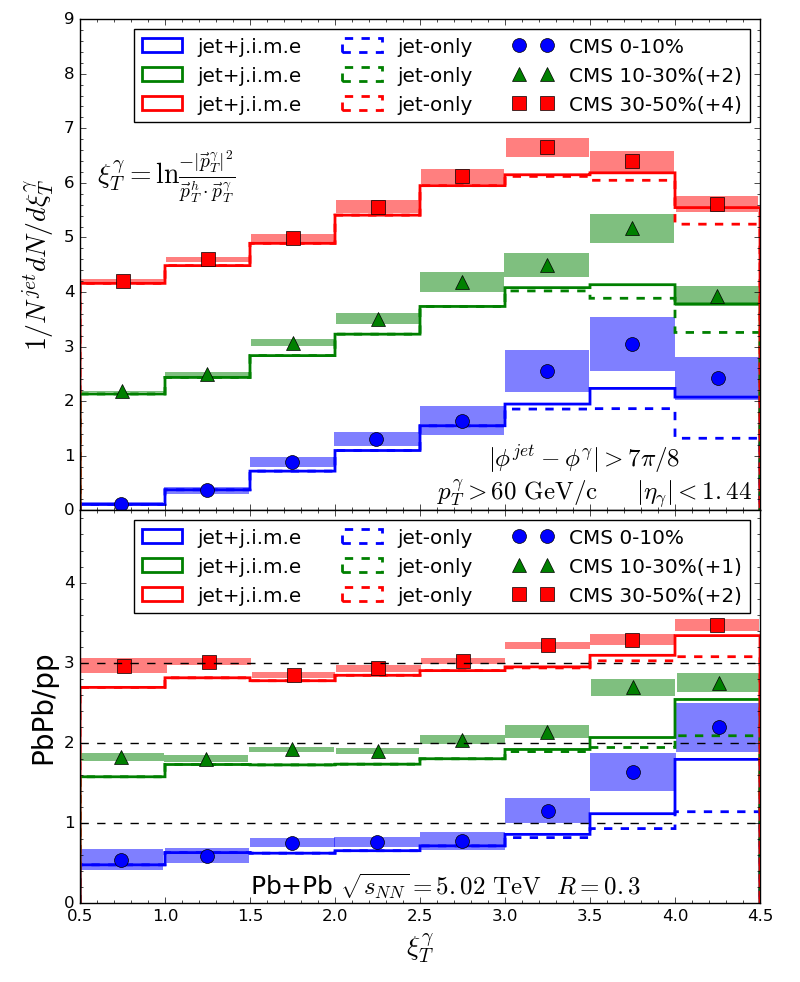} 
\caption{$\gamma$-jet fragmentation function as a function of $\xi^{\gamma}_{T}$ in Pb+Pb collisions at $\sqrt{s_{NN}}$=5.02 TeV for different centrality classes (upper panel) and the corresponding ratio of the Pb+Pb to p+p results (lower panel) as compared to CMS data~\cite{Sirunyan:2018qec}. The solid (dashed) histograms are CoLBT-hydro results with (without) jet-induced medium excitations.} 
\label{fig-ratio-gamma}
\end{figure}

Shown in Fig.~\ref{fig-ratio-jet} are CoLBT-hydro results for the $\gamma$-jet fragmentation function as a function of $\xi^{\rm jet}$ in different centralities  (0-10\%, 10-30\% and 30-50\%) of Pb+Pb collisions at $\sqrt{s_{\rm NN}}$=5.02 TeV and the corresponding ratios of the fragmentation functions in Pb+Pb to that in p+p collisions.  Charged hadrons are required to have a minimum transverse momentum $p_{T}^h>1$ GeV/$c$. The calculated $\gamma$-jet fragmentation function is normalized by the total number of photon-jet pairs $N^{\rm jet}_\gamma$ satisfying the kinematic cuts imposed by the experiment. CoLBT-hydro (solid histograms) can describe well the overall features of the medium modification of the $\gamma$-jet fragmentation function as observed in the CMS data \cite{Sirunyan:2018qec}. There is a significant enhancement of soft hadrons at large $\xi^{\rm jet}$ and slight suppression or no modification of leading hadrons at small $\xi^{\rm jet}$.  The enhancement of soft hadrons above $\xi^{\rm jet}>2.5$ is mainly due to the contribution from jet-induced medium excitation as compared to the CoLBT-hydro results without the jet-induced medium excitation (dashed histograms).  This enhancement increases from peripheral to central collisions according to CoLBT-hydro results. This centrality dependence is, however, not statistically clear in the CMS data within the experimental errors. The modification of the fragmentation function at small $\xi^{\rm jet}$ is very small or close to nonexistence. This is because of the trigger bias in the calculation of the momentum fraction $z_{\rm jet}$ ($\xi^{\rm jet}=\ln(1/z_{\rm jet})$) for fixed jet transverse momentum $p_T^{\rm jet}$. The transport of soft partons to the outer side of the jet cone will lead to the dominance of the leading hadrons in the jet finding algorithm, even though the leading parton and jet both lose energy during the jet propagation through the medium.

The energy loss of the leading jet partons will be better illustrated in the medium modification of the $\gamma$-jet fragmentation function as a function of $\xi^{\gamma}_{T}=\ln(1/z_\gamma)$  in which the momentum fractions of the final hadrons are defined relative to the transverse momentum of the direct photon $p_T^\gamma$ regardless of the final transverse momentum $p_T^{\rm jet}$ of the reconstructed jet. In this case, energy loss of the leading jet partons will lead to a strong suppression of the jet fragmentation functions at small $\xi^{\gamma}_{T}$ or large $z_\gamma$  as shown in Fig.~\ref{fig-ratio-gamma} where CoLBT-hydro results for the $\gamma$-jet fragmentation functions as a function of $\xi^{\gamma}_{T}$ in Pb+Pb collisions at $\sqrt{s_{\rm NN}}$=5.02 TeV and their ratios to that in p+p collisions are compared with CMS experimental data \cite{Sirunyan:2018qec}. We can see that both CoLBT-hydro results and CMS data show strong suppression of the jet fragmentation function at small $\xi^{\gamma}_{T}$.  Similarly, we also see a strong enhancement of low $p_{T}$ hadrons at large $\xi^{\gamma}_{T}$ while the CoLBT-hydro results without jet-induced medium excitation (dashed histograms) show little enhancement.  The magnitudes of suppression at small $\xi^{\gamma}_{T}$ due to parton energy loss and the enhancement at large $\xi^{\gamma}_{T}$ due to contributions from jet-induced medium excitation both increase from peripheral to central collisions due to the increase of average medium parton density and propagation length.


We also observe that the values of $\xi^{\rm jet}$ and $\xi^{\gamma}_{T}$ at the onset of the soft hadron enhancement depend slightly on the collision centrality and the corresponding $p_{T}^h$ is in the range $2<p_{T}^h<3$ GeV/$c$, which is consistent with the conclusion we obtained from the study of $\gamma$-hadron correlation in Au+Au collisions at the RHIC energy.  The value of $p_T^h$ at the onset of the soft hadron enhancement should reflect the average thermal energy of hadrons from the jet-induced medium excitation and therefore should be independent of the initial jet energy. This is why $\xi^{\gamma}_{T}$ of the onset is larger than  $\xi^{\rm jet}$  in Fig.~\ref{fig-ratio-jet}  because the $p_T^\gamma>60$ GeV/$c$ of the trigger photon is larger than that of the final jet $p_T^{\rm jet}>30$ GeV/$c$ on the average according to the kinematic selection of the events in both CoLBT-hydro simulations and CMS experiment. One should be able to verify this further by varying the transverse momentum of the trigger photon or the final reconstructed jets in future experimental measurements.

The enhancement of the fragmentation function at large $\xi^{\gamma}_T$  is seen to be more pronounced than that at large $\xi^{\rm jet}$ as shown in Fig.~\ref{fig-ratio-jet}. This can be understood as another trigger bias effect.  By selecting events with given $p_T^{\rm jet}$ as in the calculation of the fragmentation function in $\xi^{\rm jet}$, the selected jets are biased toward those that are initially produced close to the surface of the QGP medium. The net energy loss of the jet and corresponding jet-induced medium excitation are therefore smaller than that in events without restriction on the final jet energy as in the calculation of the fragmentation function in  $\xi^{\gamma}_T$.


\begin{figure}
\centering
\includegraphics[width=3.5in]{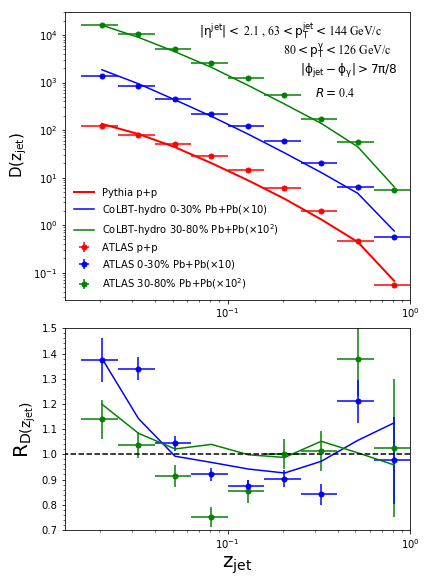} 

\caption{Fragmentation function (FF) in $\gamma$-tagged jets in p+p and Pb+Pb events in different centralities (0-30\% and 30-80\%) as a function of charged-particle longitudinal momentum fraction $z$ (upper panel) and the corresponding ratio of the Pb+Pb to p+p results (lower panel) as compared to ATLAS data \cite{Aaboud:2019oac}. }
\label{fig-atlas}
\end{figure}

To examine the kinematic dependence of the medium modification of jet fragmentation functions, we also carried out CoLBT-hydro simulations of $\gamma$-jet events in two different centrality  (0-30\% and 30-80\%) bins of Pb+Pb collisions at $\sqrt{s_{\rm NN}}$=5.02 TeV according to the ATLAS experimental analysis \cite{Aaboud:2019oac}.
In these simulations, jet-cone size is set to $R=0.4$ in the jet reconstruction using anti-$k_T$ algorithm. The transverse momentum of the trigger photon is  80$<p^{\gamma}_T<$126 GeV/c while the final reconstructed jets are restricted to 63$<p^{\rm jet}_{T}<$144 GeV/c in pseudo-rapidity range $|\eta^{\rm jet}|<$ 2.1 and the azimuthal angle $|\phi_{\rm jet}- \phi_{\gamma}|> 7\pi/8$.

Shown in Fig.~\ref{fig-atlas} are CoLBT-hydro results for $\gamma$-jet charged fragmentation functions as a function of the momentum fraction $z_{\rm jet}$ with the above kinematic constraints in p+p and Pb+Pb collisions (in two centrality bins 0-30\% and 30-80\%) at $\sqrt{s_{\rm NN}}$=5.02 TeV and their ratios as compared to ATLAS experimental data \cite{Aaboud:2019oac}.  The medium modification of the $\gamma$-jet fragmentation functions in $z_{\rm jet}$ with the ATLAS  kinematic cuts are distinctly different from that 
with the CMS experimental cuts in Fig.~\ref{fig-ratio-jet}.  While the average final jet energy in CMS analysis is smaller than the trigger photon, it is closer to or sometimes larger than that of the trigger photon in the ATLAS analysis. As a result, the fragmentation functions in the ATLAS kinematic range fall off more rapidly at large $z_{\rm jet}$. Consequently, the trigger bias effect for the leading hadrons in the fragmentation function in $z_{\rm jet}$ is stronger and leads even to an enhancement at large $z_{\rm jet}$ due to medium modification in the  central Pb+Pb collisions. With larger $p_T^\gamma$ for the trigger photons and $p_T^{\rm jet}$ for the final jets in the ATLAS analysis, the fraction of quark jets in the $\gamma$-jet events is larger than that in the CMS kinematic range.  The medium modification of the relative ratio between quark and gluon jet yield at high transverse momentum due to flavor dependence of  jet energy loss \cite{He:2018xjv} is another reason for the enhancement of the jet fragmentation function at large $z_{\rm jet}$ in Pb+Pb relative to p+p collisions. This might also be the reason for a modest suppression of the fragmentation function in the intermediate $z_{\rm jet}$ region. 

 
 The degree of the suppression and enhancement in the (30-80\%) peripheral Pb+Pb collisions is smaller than that in the central Pb+Pb collisions according to CoLBT-hydro simulations due to the shorter effective path length and in-medium effective temperature experienced by hard partons.  We do not observe statistically important modification at intermediate and large $z_{\rm jet}$ in the (30-80\%) peripheral Pb+Pb collisions. This is in contrast to the ATLAS data which show similar or more significant medium modification at intermediate and large $z_{\rm jet}$  than that in the central Pb+Pb collisions. Understanding this aspect of ATLAS data on the centrality dependence need further investigation.

\section{ summary }
The CoLBT-hydro model has been developed to simultaneously describe the transport of hard partons and the space-time evolution of the QGP medium, including  jet-induced medium excitation,  by solving the hydrodynamic equations coupled with the LBT jet transport model with a source term to account for the parton energy lost to the medium. 
We carry out CoLBT-hydro simulations of $\gamma$-jet production in Pb+Pb collisions at the LHC energy in which the final reconstructed jets  contain both hard jet shower and recoil medium partons and particles from jet-induced medium excitation. The CoLBT-hydro model is shown to provide a good prediction of medium modifications of the $\gamma$-jet fragmentation functions in Pb+Pb collisions at 5.02 TeV with different centralities as a function of two different variables $\xi^{\rm jet}$ and $\xi^{\gamma}_{T}$. 

We show that soft hadrons from jet-induced medium response lead to enhancement of fragmentation functions at both large $\xi^{\rm jet}$ and $\xi^{\gamma}_{T}$. The onset of the enhancement occurs at a constant transverse momentum $p_T^h\sim 2-3$ GeV/$c$ reflecting the thermal nature of the jet-induced medium excitation. The corresponding values of $\xi^{\rm jet}$ and $\xi^{\gamma}_{T}$ will increase with the increase of the transverse momentum of jet $p_T^{\rm jet}$ or trigger photon $p_T^\gamma$. Parton energy loss of the leading jet shower partons in medium leads to the suppression of the jet fragmentation at small $\xi^{\gamma}_{T}$ (large $z_\gamma$). The jet fragmentation functions at small $\xi^{\rm jet}_{T}$ (large $z_{\rm jet}$), however, show little medium modification and even enhancement when $p_T^{\rm jet}$ is comparable or larger than $p_T^\gamma$ due to trigger bias and medium modification of the quark to gluon jet fraction, which also leads to a modest suppression of the fragmentation function at intermediate $z_{\rm jet}$.
We have also shown the centrality dependence of the medium modification of the fragmentation function which decreases from central to peripheral collisions. However, the ATLAS data on the centrality dependence at intermediate and large $z_{\rm jet}$ need further investigation.

As we have shown in our earlier study of $\gamma$-hadron correlation at the RHIC energy, the CoLBT-hydro results on hadron spectra from jet-induced medium excitation have a weak dependence on the shear viscosity of the QGP medium. This is also true for the jet fragmentation functions at larger $\xi^{\rm jet}$ and $\xi^{\gamma}_{T}$ where jet-induced medium excitation dominates.  We will study this dependence on the shear viscosity in detail in the future, in particular the medium modification of the jet transverse profile which should be more sensitive to the transport properties of the bulk medium.

\begin{acknowledgments}

We would like to thank Yayun He and Zhong Yang for helpful discussions. This work was supported in part by the Director, Office of Energy Research, Office of High Energy and Nuclear Physics, Division of Nuclear Physics, of the U.S. Department of Energy under grant Nos. DE-SC0013460 and DE-AC02-05CH11231, the National Science Foundation (NSF) under grant Nos. ACI-1550300 and ACI-1550228 within the framework of the JETSCAPE Collaboration, and the National Natural Science Foundation of China (NSFC) under grant Nos. 11935007, 11861131009 and 11890714.

\end{acknowledgments}

\bibliography{ref_cw}

\end{document}